# MUFold-SS: Protein Secondary Structure Prediction Using Deep Inception-Inside-Inception Networks

Chao Fang[1], Yi Shang[1, *] and Dong Xu[1,2, *]

[1]Department of Electrical Engineering and Computer Science and [2]Bond Life Sciences Center, University of Missouri, Columbia, MO 65211, USA

*To whom correspondence should be addressed.

## Abstract

**Motivation:** Protein secondary structure prediction can provide important information for protein 3D structure prediction and protein functions. Deep learning, which has been successfully applied to various research fields such as image classification and voice recognition, provides a new opportunity to significantly improve the secondary structure prediction accuracy. Although several deep-learning methods have been developed for secondary structure prediction, there is room for improvement. MUFold-SS was developed to address these issues.
**Results:** Here, a very deep neural network, the deep inception-inside-inception networks (Deep3I), is proposed for protein secondary structure prediction and a software tool was implemented using this network. This network takes two inputs: a protein sequence and a profile generated by PSI-BLAST. The output is the predicted eight states (Q8) or three states (Q3) of secondary structures. The proposed Deep3I not only achieves the state-of-the-art performance but also runs faster than other tools. Deep3I achieves Q3 82.8% and Q8 71.1% accuracies on the CB513 benchmark.
**Contact:** cf797@mail.missouri.edu, shangy@missouri.edu, xudong@missouri.edu

## 1 Introduction

Protein tertiary structure prediction from amino acid sequence is a very challenging problem in computational biology (Yaseen and Li, 2014; Dill and MacCallum, 2012). However, if a protein secondary structure can be predicted accurately, it can provide useful constraints for 3D protein structure prediction. Protein secondary structure can also help identify the protein function domains and may guide the rational design of site-specific mutation experiments (Drozdetskiy *et al.*, 2015). Hence, accurate protein secondary structure prediction can help improve protein 3D structure prediction (Zhou and Troyanskaya, 2014). Pauling *et al.* (1951) proposed the earliest concept of protein secondary structure determining that the polypeptide backbone contains regular hydrogen-bonded geometry, forming α-helices and β-sheets. The prediction of protein secondary structure is often evaluated by the Q3 accuracy of three-class classification, i.e., helix (H), strand (E) and coil (C). In the 1980s, the Q3 accuracy was below 60% due to the lack of input features. In the 1990s, the Q3 accuracy reached above 70% because of using the protein evolutionary information in the form of position-specific score matrices. Since then, the Q3 accuracy has gradually improved to above 80%. However, Q8 accuracy was low until the last few years. The Q8 accuracy is another evaluation metric to evaluate the accuracy of eight-class classification: $3_{10}$-helix (G), α-helix (H), π-helix (I), β-strand (E), β-bridge (B), β-turn (T), bend (S) and loop or irregular (L) (Yaseen and Li, 2014; Zhou and Troyanskaya, 2014).

Recently, the use of deep neural networks proved to be effective and significantly improved previous accuracy on the eight-class secondary structure prediction problem. For example, Wang *et al.* (2016) applied deep convolutional neural network (CNN) with a conditional random field for secondary structure prediction. They achieved 68.3% Q8 accuracy and 82.3% Q3 accuracy on the benchmark CB513 data set. Li and Yu (2016) used a multi-scale convolutional layer followed by three stacked bidirectional recurrent layers to achieve 69.7% Q8 accuracy on the same test data set. Busia and Jaitly (2017) used CNN and next-step conditioning to achieve 71.4% Q8 accuracy on the same test data set. The current work improves upon the test accuracy of these models using different neural networks and deep architectural innovations, including residual network (He *et al.*, 2016; He *et al.*, 2016), inception network (Ioffe and Szegedy, 2015; Szegedy *et al.*, 2016; Szegedy *et al.*, 2015), batch normalization (Ioffe and Szegedy, 2015), dropout and weight constraining (Srivastava *et al.*, 2014), etc. Other than the prediction accuracy, secondary structure prediction provides an ideal testbed for exploring and testing these state-of-the-art deep-learning methods, in a similar spirit of ImageNet (http://www.image-net.org) for deep-learning method development.

The contributions of this work are: (1) Experimental results on the public CB513, CASP10, CASP11, CASP12 benchmarks show that the proposed Deep3I model outperforms existing methods (2) An open-source standalone tool, MUFold-SS, was developed and is freely available for community use. This tool can predict the protein secondary structure fast and accurately. (3) Several of the latest deep-learning methods on biological sequences were applied for the first time, which may provide useful information for applying these tools on other bioinformatics problems.

## 2 Methods

### 2.1 Inception Module

Figure 1 shows the basic Inception (Szegedy *et al.*, 2016) module of the Deep3I model. Since the convolutions with large spatial filters are computationally expensive, a hierarchical layer of convolutions with small spatial filters was used.

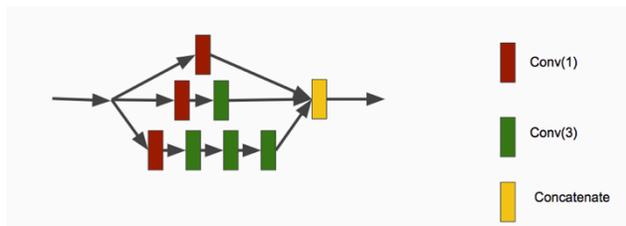

**Fig. 1.** Inception module, the red square "Conv(1)" stands for convolution operation using window size 1 and the number of filters is 100. The green square "Conv(3)" stands for convolution operation using window size 3, and the number of filters is 100. The yellow square "Concatenate" stands for feature concatenation.

### 2.2 Inception-Inside-Inception Module

Figure 2 shows how an inception-inside-inception module consists of many inception modules. Each layer in the inception module consists of an inception unit module, as a recursion of applying the inception unit inside another inception block.

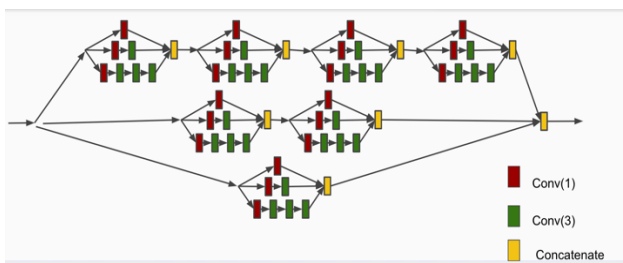

**Fig. 2.** Deep inception-inside-inception (Deep3I) module

### 2.3 Deep Inception-Inside-Inception(3I) Network

Figure 3 presents the design of Deep3I network, after many trials. Adding more Deep3I blocks is possible but requires more memory and computing time. Other Inception (Szegety *et al*., 2016) blocks were explored too.

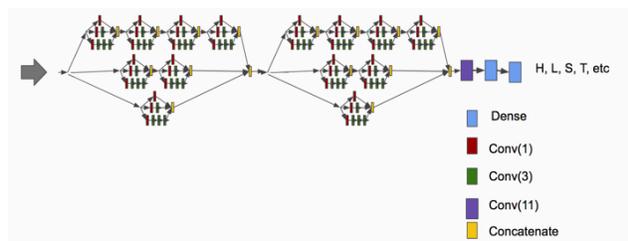

**Fig. 3.** Deep inception-inside-inception (Deep3I) network, the Deep3I network consists of two Deep3I modules, followed by a convolutional layer and a dense layer. The input of the network is the sequence and profile; the output is the predicted secondary structure labels.

### 2.4 Struct2Struct Network

The struct2struct network was proposed by (Rost and Sander, 1993). Adding the struct2struct network after the Deep3I network can fine-tune the predicted results as it takes into consideration the consecutive patterns. For example, an α-helix should consist of at least three consecutive amino acids. The predicted secondary structure from the Deep3I may violate such a pattern, i.e., not protein-like. By further feeding the initial predicted secondary structures into the second struct2struct network, the result will be fine-tuned and more protein-like. The input for the struct2struct network is the output from the previous Deep3I prediction, i.e., the predicted probability of each class from the last Softmax layer, and the output of struct2struct network is again the secondary structure label. The traditional struct2struct network described in (Rost and Sander, 1993) used two layers of simple neural networks. In this implementation, two layers of convolutional layers were used and the convolution window size was 11. The added struct2struct network may not improve the Q3(Q8) much, but it can make the prediction more protein like and help predict some small classes like B, G, S, T.

### 2.5 Batch Normalization and Dropout

TensorFlow 1.0 (Abadi *et al*., 2016) and Keras 2.0 (Chollet, 2015; https://github.com/fchollet/keras) were used for the training of the deep learning. Each convolution layer consists of four consecutive operations: 1) The convolution operation with certain kernel size. 2) The batch normalization (Ioffe and Szegedy, 2015) operation is applied to help speed up the training process and acts as a regularizer. 3) The activation operation, 'ReLU,' (Radford *et al*., 2015) was used as an activation function. 4) The dropout (Srivastava *et al*., 2014) operation to prevent the neural network from overfitting randomly drops neurons during the deep network training process such that the network can avoid too much co-adapting.

During implementation, the dropout rate was set to 0.4. During network training, the learning rate scheduler from Keras was also used to control the learning rate. The early stopping mechanism from Keras was used to stop network training when the monitored validation quantity (such as validation loss and/or validation accuracy) stopped improving. The "patience" (i.e., the number of epochs with no improvement after which training was stopped) was set between 8 to 12 in experiments. TensorBoard from Keras was used to visualize dynamic graphs of the training and validation metrics.

### 2.6 Input Features

The input features we used are the same as (Busia and Jaitly, 2017). The amino acid at position *i* is represented as one-hot vector. There are 20 different types of amino acids. Some special amino acid cases were handled as follows: Amino Acid '*X*' was treated as amino acid '*A*'. '*B*' was treated as amino acid '*N*'. '*Z*' was treated as amino acid '*Q*'. Since the input size was fixed at 700, if the input protein sequences were less than 700 amino acids; the remaining amino acid positions were padded with the '*NoSeq*' label, which means no amino acid was there. Current implementation can handle any protein sequence length less or equal to 700 amino acids. The protein sequences will be split into smaller segments if they have more than 700 amino acids. A protein sequence is represented as a 700-by-21 array in the system. Next, protein profiles were generated using PSI-BLAST (Altschul *et al*., 1997), as performed in some previous work (Wang et al., 2016; McGuffin *et al*., 2000). The PSI-BLAST tool parameters were set as follows: *evalue*: 0.001, *num_iterations*: 3, *db*, and UniRef50 to generate a position-specific scoring matrix (PSSM). PSSM is then transformed by the sigmoid function so that the value is in range (0,1). The profile information is also represented as 700-by-21 array. Hence, the input dimension is 700-by-42 in total.

### 2.7 Proposed Network is Different from Others

The proposed Deep3I network (see Fig. 2) differs from the previous network (Li and Yu, 2016; Busia and Jaitly, 2017) in that the latter ones used residual blocks and multi-scale layer containing CNN layers with a convolution window size of 3, 7, and 9 to discover protein local and global context. Deep3I consists of stacked CNN layers, whose convolution window size is only 3. When stacked deep convolution blocks are put together, they can perform both local and global context extraction. Applying convolution on top of convolution, the sliding window will cover a wide range of protein sequences by using this hierarchical convolutional operation. Besides that, a struct2struct (Rost and Sander, 1993) network is added after Deep3I for fine-tuning the secondary structure prediction. This network acts as a fine-tuning layer on top of Deep3I and can further make the prediction results more protein-like.

### 2.8 Data Sets
In this work, four public data sets were used:
1) **CullPDB** (Wang and Dunbrack, 2003) used in Zhou and Troyanskaya (2014) and Busia and Jaitly (2017). The CullPDB data set from Zhou and Troyanskaya, (2014) was constructed before January 2014, and any two proteins in this set shared less than 25% sequence identity with each other. This CullPDB contained 6128 proteins. The filtered data set of this CullPDB had a sequence identity of less than 25% with the CB513 test data, and it contained 5534 protein sequence after filtering. Finally, 5278 protein sequences were randomly chosen from the filtered CullPDB to form a training set, and the remaining 256 sequences formed the validation set.
2) **CB513 benchmark** used in Zhou and Troyanskaya (2014), Wang et al. (2016), Li and Yu (2016), and Busia and Jaitly (2017). This benchmark was widely used and was chosen from secondary structure tools for performance comparison.
3) **JPRED** (Drozdetskiy *et al.*, 2015) data set. This benchmark contains non-redundant proteins for training and testing, and each of the protein belongs to a different superfamily.
4) **CASPs 10, 11 and 12.** Critical assessment of protein structure prediction (CASP) is a community-wide protein structure prediction biannual competition. CASP targets were widely used for benchmark protein prediction tools.

The data set (1) was used to train the deep network while data sets (2–4) were only used for testing and comparison with other state-of-the-art tools.

### 2.9 Performance Evaluation Metric
Q3 and Q8 were used as a performance metric, as commonly used in Zhou and Troyanskaya (2014), Wang et al. (2016), Li and Yu (2016), and Busia and Jaitly (2017). The Q3 or Q8 accuracy measured the percentage of residues being correctly predicted among three-state or eight-state protein secondary structure classes. Besides Q3 and Q8, Matthews correlation coefficient (Matthews, 1975) was also used as performance metric, as it takes true and false positives and negatives into account and provides a more balanced measurement of quality network classification capability.

## 3 Results and Discussion
The following experiments shows: (1) Deep3I can perform the state-of-the-art Q3 (Wang *et al.*, 2016) and Q8 (Busia and Jaitly, 2017), and (2) profile search time is relatively shorter than current available tools because we used a filtered version of the UniRef50 database.

### 3.1 How Does Database Size Affect PSI-BLAST Search Time and Q3 Accuracy?

The PSI-BLAST can generate the protein sequence profile in a short time if the search database is small, while a larger search database takes a longer time to get the profile and the prediction accuracy may not always increase. To verify this, we designed and performed the following experiments: The JPRED data set was used. JPRED itself contains 1348 sequences as training set and 149 protein sequences as test sets. Four different databases were downloaded from UniProt (http://www.uniprot.org/downloads). They are Swiss_Prot (0.08GB), UniRef50 (4.3GB), UniRef90 (12GB) and UniRef100 (25GB). Take the UniRef50 database as an example: This database was used to perform a PSI-BLAST search with the parameter setup of *evalue* 0.001 and *num_iterations* 3 on all protein sequences in the JPRED data set. Then, the training set was used to train the Deep3I network, and the test set was used to report Q3, as shown in Table 1. The UniRef50 yielded the best results as a larger database was not needed to yield a better performance.

**Table 1.** Comparison of the profile generation execution time and Deep3I Q3 accuracy using different databases

| Database | Zipped file Size (GB) | Profile average generation time | Q3 % |
| --- | --- | --- | --- |
| Swiss_Prot* | 0.08 | 7.2s | - |
| UniRef50 | 4.3 | 6.6m(+/-5.4m) | 82.61(+/-0.42) |
| UniRef90 | 12 | 36.6m(+/-20.4m) | 81.74(+/-0.5) |
| UniRef100 | 25 | 1.06h(+/-34.8m) | 79.61(+/-0.4) |

*The Swiss_Prot database is too small; thus, many protein profiles are not generated.

### 3.2 How does number of PSI-BLAST iterations affect Q3?
Same JPRED data set were used. The PSI-BLAST search database is UniRef50 and *evalue* is set to 0.001. Four experiments were performed with different *num_iteration* of PSI-BLAST ranging from 2 to 5. Table 2 shows that too few or too many PSI-BLAST iterations do not yield good profile. The number of iterations of PSI-BLAST should be set to 3.

**Table 2.** Comparison of the profile generation execution time Deep3I Q3 accuracy using different iteration of PSI-BLAST

| |S| | # of PSI-BLAST iterations | Timing (m) | Q3 % |
| --- | --- | --- | --- |
| UniRef50 8.7GB | 2 | 5.69(+/-2.58) | 82.17(+/-0.31) |
| | 3 | 6.42(+/-5.47) | 82.61(+/-0.42) |
| | 4 | 9.35(+/-5.3) | 82.28(+/-0.16) |
| | 5 | 11.48(+/-6.25) | 81.97(+/-0.21) |

PSI-BLAST evalue is set to 0.001 for all experiments.

**Table 3.** PSI-BLAST database comparison

| Database | Size(GB) | # protein sequence | Max sequence length |
| --- | --- | --- | --- |
| 1.UniRef50 | ~8.7 | 21,859,863 | 36,805 |
| 2.UniRef50_shrunk | ~8.1 | 19,430,324 | 70~3,000 |
| 3.UniRef50_smaller | ~7.0 | 18,691,241 | 70~1,000 |
| 4.UniRef50_from_sspro | ~5.3 | * | * |
| 5.UniRef50_evenSmall | ~5.1 | 16,115,059 | 70~500 |
| 6.UniRef50_tiny | ~2.7 | 9,634,815 | 100~300 |

The UniRef50 database was downloaded on 2017/4/12.

*The SSpro package contains a small UniRef50 database produced several years ago. The Fasta file was not included in the package and the specific number of the protein sequences and the maximum sequence length are not available.

### 3.3 Could the Database be Even Smaller than UniRef50?

Based on the information presented thus far, one could assume that the smaller the database, the better the Q3. One might ask: What will happen if an even smaller database is used for a PSI-BLAST search? To find out, we filtered a UniRef50 data set and kept those protein sequences whose length fell between 70 and 3000 thereby forming the shrunk database, UniRef50_shrunk. Some other smaller databases were built, as shown in Table 3. Table 4 shows that the UniRef50_smaller was sufficient for PSI-BLAST to use as a database and saves computing time. But when the database became even smaller, the prediction accuracy started to drop or fail. Another option would be to apply CD-Hit to shrink the database with a lower sequence similarity threshold.

**Table 4.** Performance of SSpro, PSI-PRED and MUFold-SS with varying database sizes on an average PSI-BLAST profile search with CPU time, Q3 and Q8 using the CB513 benchmark

|  | BLAST version | Database abbreviation | CPU execution time(m) | Q3 % | Q8 |
|---|---|---|---|---|---|
| SSpro w/o template | 2.2.26 | 4.SSpro's | 6.86 | 78.6 | 66.5 |
| PSIPRED | 2.2.26 | 4.SSpro's | 6.28 | 79.2 | N/A |
| MUFold-SS | 2.6.0 | 1.UniRef50 | 8.23 | 82.98 | 71.05 |
|  |  | 2.shrunk | 6.86 | 82.92 | 71.03 |
|  |  | 3.smaller | 5.51 | 82.83 | 71.09 |
|  |  | 4.SSpro's | 5.28 | 82.67 | 70.51 |
|  |  | 5.evenSmaller | 3.57 | 82.44 | 70.39 |
|  |  | 6.tiny | 1.71 | Fail, too many profiles—no hit | |
|  | 2.2.26 | 2.shrunk | 6.02 | 81.58 | 69.11 |
|  |  | 3.smaller | 5.02 | 81.68 | 69.19 |
|  |  | 4.SSpro's | 4.03 | 81.49 | 68.99 |

The database was built on 2017/4/12, 8.1GB. PSI-BLAST evalue is set to 0.001.

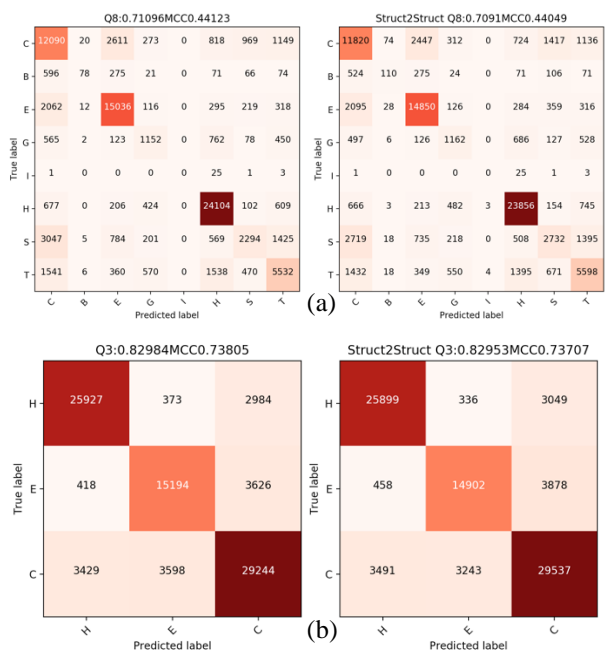

**Fig. 4.** Effect of (a) Q8 and (b) Q3 using struct2struct network. The left confusion matrix shows the prediction result by Deep3I, and the right confusion matrix shows the prediction result by adding another two-layer neural network serving as struct2struct network.

### 3.4 Struct2Struct Network

The struct2struct network improved the Q3(Q8) performance but not that significantly. Since the most improvement can be achieved by the Deep3I network, the struct2struct can act as a refinement layer (See Fig. 4a for Q8 and Fig. 4b for Q3). Nevertheless, it correctly predicted more small classes like B, G, S and T for Q8.

### 3.5 Benchmark MUFold-SS against other state-of-the-art tools in terms of time, Q3 and Q8

This experiment was developed to benchmark MUFold-SS against current state-of-the-art tools. Two widely used tools were selected: SCRATCH_1D (SSPro ab initio) (Cheng *et al.*, 2005, Magnan and Baldi, 2014) and PSIPRED (McGuffin, 2000). The benchmark CB513 data set reported performance. For the PSI-BLAST database selection, to be fair for all tools, a smaller version of UniRef50, which is included in the SSPro package, was used to generate PSI-BLAST profiles. Also for PSI-BLAST version, SSPro included the legacy BLAST 2.2.26. PSIPRED does not have a PSI-BLAST package included, so legacy BLAST 2.2.26 was installed as its replacement. Deep3I was developed using the latest BLAST+ package 2.6.0+. Three results were reported: average program CPU time (minutes/sample), Q3 and Q8. As shown in Table 5, MUFold-SS can predict the secondary structure in a fast and more accurate way compared with other methods.

**Table 5.** Benchmark MUFold-SS against SSPro, PSIPRED in terms of time, Q3 (%) and Q8 (%) using CB513 data set

|  | CPU Execution time | Q3 | Q8 |
|---|---|---|---|
| SSPro w/o template | 6.86 m | 78.6 | 66.5 |
| PSIPRED | 6.28 m | 79.2 | N/A |
| MUFold-SS | 5.28 m | **82.6** | **70.7** |

### 3.6 Benchmark MUFold-SS against State-of-the-Art Tools Using CASP 10, 11, and 12 data sets

Each protein from the CASP data set was stored in the PDB format. To prepare the data set, the PDB files were downloaded from the official CASP website under the target directory: http://prediction-center.org/download_area/CASP10/targets/. Note that some of the PDB files provided do not cover the corresponding FASTA file. For example, the T0644.fasta contains 166 amino acids; however, the T0644.pdb contains only 141 amino acids. To be more consistent, the extracted sequence from PDB provided by CASP was used. The DSSP program (Touw *et al.*, 2015; Kabsch and Sander, 1983) was used to get the secondary structure label from the PDB files. Some of the PDB files (T0675, T0677 and T0754) could not generate the DSSP results, and they were discarded. Some protein sequences (T0709, T0711, T0816 and T0820) are too short and PSI-BLAST did not have a hit; hence, they could not be considered in the evaluation, either. Overall, the effective proteins used are CASP10 (98 out of 103), CASP 11 (83 out of 85) and CASP 12 (40 out of 40). The Q3 and Q8 results are shown in Table 6 and Table 7, respectively. Note that only one number is reported for each Q3(Q8) in CASPs 10, 11, and 12.

**Table 6.** Q3(%) comparison between MUFold-SS with other state-of-the-art methods.

|  | CB513 | CASP10 | CASP11 | CASP12 |
| --- | --- | --- | --- | --- |
| SSpro (w/o template)* | 78.5 | 78.5 | 77.6 | 76.0** |
| SSpro (w/ template)* | **90.7** | 84.2 | 78.4 | 76.6** |
| SPINE-X* | 78.9 | 80.7 | 79.3 | - |
| PSIPRED* | 79.2 | 81.2 | 80.7 | 80.4** |
| JPRED* | 81.7 | 81.6 | 80.4 | - |
| RaptorX-SS8* | 78.3 | 78.9 | 79.1 | - |
| DeepCNF-SS* | 82.3 | 84.4 | **84.7** | - |
| MUFold-SS** | 82.98 | **85.3** | 83.4 | **80.8**** |

\* Results reported by (Wang et al., 2016)
\*\* Experiment results conducted by authors

**Table 7.** Q8 (%) comparison between DeepCNF-SS and MUFold-SS

|  | CB513 | CASP10 | CASP11 | CASP12 |
| --- | --- | --- | --- | --- |
| SSpro(w/o template)* | 63.5 | 64.9 | 65.6 | 63.1** |
| SSpro (w/ template) * | **89.9** | 75.9 | 66.7 | 64.1** |
| ICML2014#* | 66.4 | - | - | - |
| RaptorX-SS8* | 64.9 | 64.8 | 65.1 | - |
| DeepCNF-SS* | 68.3 | 71.8 | 72.3 | - |
| MUFold-SS** | 71.05 | **75.9** | **72.9** | **69.9**** |

\* Results reported by (Wang et al., 2016)
\*\* Experiment results conducted by authors
#ICML2014 (Zhou and Troyanskaya, 2014)

## 4 Conclusion and Future Work

In this work, a new deep neural network, Deep3I, was introduced and presented as an improved program for protein secondary structure prediction. Q3 performed statistically on par with other current state-of-the-art methods (Wang et al., 2016). Q8 was 0.7% less effective than the best available method (Busia and Jaitly, 2017), but better than all the other methods. In Tables 7 and 8, please note that the CullPDB data used to train the network is from (Zhou and Troyanskaya, 2014), which was uploaded in January 2014 and probably generated before 2014. CASP 10, 11, and 12 were downloaded on May 2012, 2014 and 2016. Better results from CASP 10 were assumed to be due to possible redundant or similar sequences found in the CullPDB data set. Likewise, the CASP12 results were slightly worse for the same tentative reason. Prediction accuracy improves greatly when used with the most updated data set of CullPDB. Compared with previous deep-learning applications in protein secondary structure prediction, our application used a more cutting-edge deep-learning architecture, adopted a struct2struct network for making the prediction results more protein-like, and fine-tuned the parameters and the search database for more accurate and faster performance. We developed the first open-source deep-learning based secondary structure prediction tool MUFold-SS. It was implemented and provided to the research community. The open source advantage adds significant value to this work, as it allows other researchers to easily apply this deep-learning framework for many other research problems.

Future work will include several areas. The protein sequence input is one-hot vector, which is a sparse matrix. In the future work, ProtVec (Asgari and Mofrad, 2015) will be explored to represent the sequence so that it will become a much denser and picture-like matrix, which may improve the Q3 and Q8 accuracy. In the PSI-BLAST iteration experiment, we found an interesting phenomenon showing that more iterations may not lead to higher accuracy. Also, a larger PSI-BLAST database does not necessary improve the result of Q3 or Q8 because the profile can diverge to biologically irrelevant hits. The reasonable iteration number should be three. Besides protein secondary structure prediction, Deep3I can also be extended to predict solvent accessibility, contact number, and protein order/disorder regions. These predicted features are also useful for protein structure prediction or quality assessment.

## Funding

This work was partially supported by the National Institutes of Health grant R01-GM100701. The high-performance computing infrastructure was partially supported by the National Science Foundation under grant number CNS-1429294.

*Conflict of Interest:* None declared.

## References

Abadi, M., Agarwal, A., Barham, P., Brevdo, E., Chen, Z., Citro, C., ... & Ghemawat, S. (2016). Tensorflow: Large-scale machine learning on heterogeneous distributed systems. arXiv preprint arXiv:1603.04467.

Altschul, S. F., Madden, T. L., Schäffer, A. A., Zhang, J., Zhang, Z., Miller, W., & Lipman, D. J. (1997). Gapped BLAST and PSI-BLAST: a new generation of protein database search programs. *Nucleic acids research*, *25*(17), 3389-3402.

Asgari, E., & Mofrad, M. R. (2015). Continuous distributed representation of biological sequences for deep proteomics and genomics. *PloS one*, *10*(11), e0141287.

Busia, A., & Jaitly, N. (2017). Next-Step Conditioned Deep Convolutional Neural Networks Improve Protein Secondary Structure Prediction. *arXiv preprint arXiv:1702.03865*

Biegert, A., & Söding, J. (2009). Sequence context-specific profiles for homology searching. *Proceedings of the National Academy of Sciences*, *106*(10), 3770-3775.

Boratyn, G. M., Schäffer, A. A., Agarwala, R., Altschul, S. F., Lipman, D. J., & Madden, T. L. (2012). Domain enhanced lookup time accelerated BLAST. *Biology direct*, *7*(1), 12.

Cheng, J., Randall, A. Z., Sweredoski, M. J., & Baldi, P. (2005). SCRATCH: a protein structure and structural feature prediction server. *Nucleic acids research*, *33*(suppl 2), W72-W76.

Chollet, F. (2015). Keras.

Dill, K. A., & MacCallum, J. L. (2012). The protein-folding problem, 50 years on. *Science*, *338*(6110), 1042-1046.

Drozdetskiy, A., Cole, C., Procter, J., & Barton, G. J. (2015). JPred4: a protein secondary structure prediction server. *Nucleic acids research*, gkv332.

He, K., Zhang, X., Ren, S., & Sun, J. (2016). Deep residual learning for image recognition. In *Proceedings of the IEEE Conference on Computer Vision and Pattern Recognition* (pp. 770-778).

He, K., Zhang, X., Ren, S., & Sun, J. (2016, October). Identity mappings in deep residual networks. In *European Conference on Computer Vision* (pp. 630-645). Springer International Publishing.

Ioffe, S., & Szegedy, C. (2015). Batch normalization: Accelerating deep network training by reducing internal covariate shift. *arXiv preprint arXiv:1502.03167*.

Kabsch, W., & Sander, C. (1983). Dictionary of protein secondary structure: pattern recognition of hydrogen-bonded and geometrical features. *Biopolymers*, 22(12), 2577-2637.

Li, Z., & Yu, Y. (2016). Protein Secondary Structure Prediction Using Cascaded Convolutional and Recurrent Neural Networks. *arXiv preprint arXiv:1604.07176*.

Magnan, C. N., & Baldi, P. (2014). SSpro/ACCpro 5: almost perfect prediction of protein secondary structure and relative solvent accessibility using profiles, machine learning and structural similarity. *Bioinformatics*, *30*(18), 2592-2597.

Matthews, B. W. (1975). Comparison of the predicted and observed secondary structure of T4 phage lysozyme. *Biochimica et Biophysica Acta (BBA)-Protein Structure*, *405*(2), 442-451.

McGuffin, L. J., Bryson, K., & Jones, D. T. (2000). The PSIPRED protein structure prediction server. *Bioinformatics*, *16*(4), 404-405.

Pauling, L., Corey, R. B., & Branson, H. R. (1951). The structure of proteins: two hydrogen-bonded helical configurations of the polypeptide chain. *Proceedings of the National Academy of Sciences*, *37*(4), 205-211.

Radford, A., Metz, L., & Chintala, S. (2015). Unsupervised representation learning with deep convolutional generative adversarial networks. *arXiv preprint arXiv:1511.06434*.

Rost, B., & Sander, C. (1993). Prediction of protein secondary structure at better than 70% accuracy. *Journal of molecular biology*, *232*(2), 584-599.

Srivastava, N., Hinton, G. E., Krizhevsky, A., Sutskever, I., & Salakhutdinov, R. (2014). Dropout: a simple way to prevent neural networks from overfitting. *Journal of Machine Learning Research*, *15*(1), 1929-1958.

Szegedy, C., Ioffe, S., Vanhoucke, V., & Alemi, A. (2016). Inception-v4, inception-resnet and the impact of residual connections on learning. *arXiv preprint arXiv:1602.07261*.

Szegedy, C., Liu, W., Jia, Y., Sermanet, P., Reed, S., Anguelov, D., ... & Rabinovich, A. (2015). Going deeper with convolutions. In *Proceedings of the IEEE Conference on Computer Vision and Pattern Recognition* (pp. 1-9).

Szegedy, C., Vanhoucke, V., Ioffe, S., Shlens, J., & Wojna, Z. (2016). Rethinking the inception architecture for computer vision. In *Proceedings of the IEEE Conference on Computer Vision and Pattern Recognition* (pp. 2818-2826).

Touw, W. G., Baakman, C., Black, J., te Beek, T. A., Krieger, E., Joosten, R. P., & Vriend, G. (2015). A series of PDB-related databanks for everyday needs. Nucleic acids research, 43(D1), D364-D368.

UniProt Consortium. (2017). UniProt: the universal protein knowledgebase. *Nucleic acids research*, *45*(D1), D158-D169.

Wang, G., & Dunbrack, R. L. (2003). PISCES: a protein sequence culling server. *Bioinformatics*, *19*(12), 1589-1591.

Wang, S., Peng, J., Ma, J., & Xu, J. (2016). Protein secondary structure prediction using deep convolutional neural fields. *Scientific reports*, *6*.

Yaseen, A., & Li, Y. (2014). Context-based features enhance protein secondary structure prediction accuracy. *Journal of chemical information and modeling*, *54*(3), 992-1002.

Zhang, Z., Miller, W., Schäffer, A. A., Madden, T. L., Lipman, D. J., Koonin, E. V., & Altschul, S. F. (1998). Protein sequence similarity searches using patterns as seeds. *Nucleic acids research*, *26*(17), 3986-3990.

Zhou, J., & Troyanskaya, O. (2014, January). Deep supervised and convolutional generative stochastic network for protein secondary structure prediction. In *International Conference on Machine Learning* (pp. 745-753).